\documentclass[epj]{svjour}

\usepackage{graphics}
\usepackage{verbatim}
\usepackage{graphicx}
\usepackage{url}
\usepackage{color}
\usepackage{amsmath}
\newcommand{\NNLOsat}{NNLO$_{\rm sat}$}
\newcommand{\NNLOgod}{$\Delta$NNLO$_{\rm GO}$(450)}

\renewcommand\refname{\vskip -1cm}

\begin{document}
\title{Measurement of the $^{40}$Ar(e,e$^{\prime}$) elastic scattering cross section with a novel gas-jet target}
%\title{Measurement of the electron-argon elastic nuclear cross section with a  jet target}
%\title{Operation of an argon gas-jet target with electron beams}

\author{
M.~Littich\inst{1} \and 
L.~Doria\inst{1}\thanks{\emph{Corresponding author,} doria@uni-mainz.de} \and 
P.~Brand\inst{2} \and
P.~Achenbach\inst{1} \and
S.~Aulenbacher\inst{1} \and
S.~Bacca\inst{1} \and
J.C.~Bernauer\inst{3} \and
M.~Biroth\inst{1} \and
D.~Bonaventura\inst{2} \and
D.~Bosnar\inst{4}\and
M.~Christmann\inst{1} \and
E.~Cline\inst{3,5} \and
A.~Denig\inst{1} \and
M.~Distler\inst{1} \and
A.~Esser\inst{1} \and
I.~Fri\v{s}\v{c}i\'{c}\inst{4} \and
J.~Geimer\inst{1} \and
P.~Gülker\inst{1}\and
M.~Hoek\inst{1}\and
P.~Klag\inst{1}\and
A.~Khoukaz\inst{2}\and
M.~Lau\ss \inst{1}\and
S.~Lunkenheimer\inst{1}\and
T.~Manoussos\inst{1}\and
D.~Markus\inst{1}\and
H.~Merkel\inst{1}\and
M.~Mihovilovi\v c\inst{6,7}\and
U.~Müller\inst{1}\and
J.~Pochodzalla\inst{1}\and
B.S.~Schlimme\inst{1}\and
C.~Sfienti\inst{1}\and
J.E.~Sobczyk\inst{1}\and
S.~Stengel\inst{1}\and
E.~Stephan\inst{8}\and
M.~Thiel\inst{1}\and
S.~Vestrick\inst{2}\and
A.~Wilczek\inst{8}\and
L.~Wilhelm\inst{1}
}                     % Do not remove
\institute{Institut f\"ur Kernphysik, Johannes Gutenberg-Universit\"at, D-55128, Mainz, Germany\and 
Institut f\"ur Kernphysik, Universität M\"unster, D-48149, M\"unster, Germany \and
Center for Frontiers in Nuclear Science, Department of Physics and Astronomy, Stony Brook University, 11794, New York, USA \and
Department of Physics, Faculty of Science, University of Zagreb, Zagreb, Croatia \and
Laboratory for Nuclear Science, Massachusetts Institute of Technology, 02139, Cambridge, USA \and
Jo\v zef Stefan Institute, SI-1000, Ljubljana, Slovenia \and
Faculty of Mathematics and Physics, University of Ljubljana, SI-1000, Ljubljana, Slovenia \and
Institute of Physics, University of Silesia in Katowice, 41-500, Chorzow, Poland
}
\date{Received: date / Revised version: date}
\abstract{
We report on a measurement of elastic electron scattering on argon performed with a novel cryogenic gas-jet target at the Mainz Microtron accelerator MAMI. The luminosity is estimated with the thermodynamical parameters of the target
and by comparison to a calculation in distorted-wave Born approximation.
The cross section, measured at new momentum transfers of 1.24~fm$^{-1}$ and 1.55~fm$^{-1}$ is in agreement with
previous experiments performed with a traditional high-pressure gas target, as well as with
modern {\em ab-initio} calculations 
employing state-of-the-art nuclear forces from chiral effective field theory. 
The nearly background-free measurement highlights the optimal properties of the gas-jet target for elements 
heavier than hydrogen, enabling new applications in hadron and nuclear physics.
%that will be addressed with the MAGIX experiment in the future. 
}

 % end of PACS codes

\maketitle

\section{Introduction}
\label{sec:Introduction}
Argon is a noble gas that plays a pivotal role in modern particle physics experiments. Its unique combination of scintillation properties, abundance, and affordability makes it an ideal detector material for a wide range of applications. In particular, liquid argon time projection chambers (LArTPCs) have emerged as a cornerstone technology in neutrino physics and rare-event searches, providing exceptional spatial and energy resolutions for tracking ionizing particles~\cite{Rubbia1977}. 
On the neutrino physics side, detectors presently operating with this technology, such as MicroBooNE~\cite{microboone}, ICARUS~\cite{ICARUS}, and SBND~\cite{SBND}, 
are paving the way to DUNE~\cite{DUNE}, a next-generation
long-baseline neutrino experiment based on kt-scale LArTPCs. On the dark matter side,  the success of DEAP~\cite{DEAP} and DarkSide-50~\cite{DarkSide50} is setting the stage for the future DarkSide-20k experiment \cite{DS20k-0,DS20k-1,DS20k-2}.

To understand the response of argon  to external probes, electron scattering experiments can be performed to gain insights into its nuclear structure properties~\cite{mariani}. Collecting this information is crucial for the above-mentioned experiments, where precise modeling of particle-argon interactions is necessary~\cite{e4nu}.
 \begin{figure*}
  \centering
  \resizebox{0.9\textwidth}{!}{%
  \includegraphics{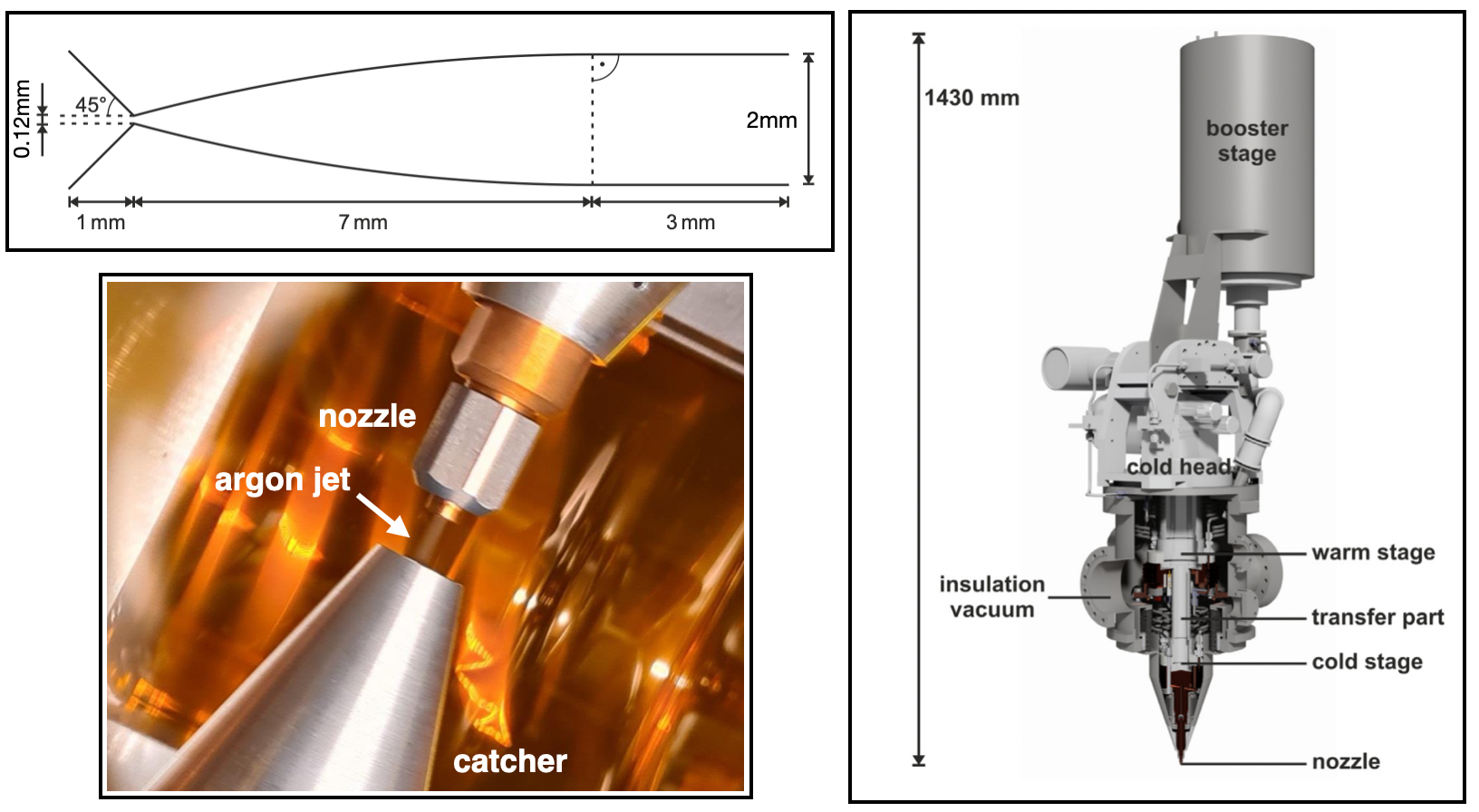}
   }
  \caption{{\bf (Top Left)} Profile of the Laval nozzle. The cooled gas flows
    from the left to the right direction. {\bf (Bottom Left)} Picture of the jet target
    in operation with argon. The jet is visible between the nozzle and the catcher.
    {\bf (Right)} Schematic view of the jet target (reproduced from Ref.~\cite{A1hydrogen}).}
  \label{fig:JTpicture}
\end{figure*}
While electron scattering has a rich history of important discoveries \cite{Hofstadter}, achieving high precision
is often limited by the thickness of the target.
Multiple scattering and energy straggling represent serious obstacles towards percent-level precision on cross section measurements. When possible, the use of gas targets can mitigate these issues. However, traditional gas targets introduce additional backgrounds  due to the presence of a containment vessel and removing their effects can be difficult. 
Therefore, to achieve high precision for argon, a windowless gas-jet target is the optimal choice. 

In this work, we present a measurement of the elastic electron-argon scattering cross section using a novel gas-jet target and the MAMI electron accelerator~\cite{MAMI1,MAMI2}. 
This experiment serves as a proof of principle for the use of the gas-jet target with elements heavier than hydrogen and demonstrates its potential for studies with the MAGIX experiment~\cite{Magix} at the future MESA facility~\cite{MESA1,MESA2}, where electron beam currents in the mA range will enable novel experiments in nuclear and hadron physics.

The cross section measurement of elastic electron scattering offers also a benchmark for state-of-the-art theoretical calculations. In particular,
{\em ab-initio} theory aims to describe nuclear systems starting from fundamental interactions between nucleons, rooted in quantum chromodynamics via frameworks such as chiral effective field theory~\cite{Epelbaum:2008ga,machleidt2011,hammer2020}. This approach provides a rigorous and systematic method for calculating nuclear properties and reaction dynamics, enabling predictions of observables like cross sections for electron and neutrino-nucleus scattering~\cite{Lovato:2017cux,Acharya:2019fij,Payne:2019wvy,Lovato_2020,Sobczyk:2021dwm,Sobczyk:2023mey}. In the context of neutrino and dark matter physics, where the interaction of particles with atomic nuclei is a critical input for experiments like DUNE~\cite{DUNE} and Hyper-Kamiokande~\cite{HyperK}, {\em ab-initio} methods are particularly interesting. They allow for the calculation of both elastic and inelastic effects with quantified uncertainties, which are essential to
advance our understanding of the complex nuclear responses.
 
Elastic electron-nucleus scattering serves as an ideal benchmark for this purpose, as it provides a clean probe of the nuclear electromagnetic response, which has a simpler theoretical structure than the weak neutrino or dark matter interactions \cite{DEAPeffective}.

Such comparisons may in principle enable the refinement of nuclear models  and improve  the treatment of nuclear correlations, which, in turn, reduce systematic uncertainties in  detector simulations. This is particularly important for next-generation neutrino experiments, where precise modeling of neutrino-nucleus interactions is critical for achieving their physics goals~\cite{coloma2013,Mihovilovic:2024ymj}.
Ultimately, by combining {\em ab-initio} theory with experimental validation, the field can progress toward a deeper and more accurate understanding of nuclear dynamics in neutrino interactions.

The paper is organized as follows: Section \ref{sec:ExperimentalSetup}
describes the experimental apparatus and the jet target. Section \ref{sec:Operation} presents the theoretical
estimate of the target areal density, and in Section \ref{sec:CrossSection} the elastic electron-argon cross section measurement is presented. Data are then compared to
a previous measurement and to {\em ab-initio} theoretical calculations based on modern chiral potentials.

%%%%%%%%%%%%%%%%%%%%%%%%%%%%%%%%%%%%%%%%%%%%%%%%%%%%%%%%%%%%%%%%%%%%%%%%%%%%%%%%%%%%
\section{Experimental Setup and the Jet Target}
\label{sec:ExperimentalSetup}
The jet target was operated at the A1 experimental setup \cite{A1spectrometers} 
which consists of three magnetic spectrometers (conventionally called A, B, and C) that can rotate around the target. 
The spectrometers have solid angle acceptances of 28~msr (for A and C) and 5.6~msr (for B),
and momentum acceptances of 20\% (for A), 15\% (for B), and 25\% (for C).
The relative momentum resolution is $\delta p/p \approx 10^{-4}$ and the angular resolution $\delta\theta \approx 3$~mrad.
The detector system is similar in all the three spectrometers and consists of vertical drift chambers
for tracking and momentum determination, a double-plane of plastic scintillator paddles for triggering,
and Cherenkov detectors for particle identification. The MAMI accelerator~\cite{MAMI1,MAMI2} provides a 100\% duty cycle beam with an energy of up to 1.6~GeV and a maximum current of 100~$\mu$A. The MAMI beam is directed at the spectrometers' geometric center of rotation, where the gas-jet target is positioned.

To prepare the target gas, the process begins with a standard argon gas bottle at 15~bar pressure which is then reduced to 2~bar by a flow controller and cooled in a booster stage using liquid nitrogen. 
The nitrogen supply is automatically maintained via readings from a level meter. From there, the gas passes through copper windings surrounding two cryogenic cold-head stages, each one controlled with heaters and temperature sensors for precise regulation.
The second stage directly determines the gas temperature at the nozzle (Figure~\ref{fig:JTpicture}, bottom left). 
The gas is accelerated through a Laval-type nozzle (Figure~\ref{fig:JTpicture}, top left), forming a supersonic jet beam. 
To limit thermal losses, both the booster and cold-head stages are housed in separate vacuum chambers, minimizing interaction with the environment.
Beneath the nozzle, a catcher, connected to a dedicated vacuum system, collects the gas while maintaining a stable vacuum in the target vessel. For further technical details and for the results of testing with hydrogen, we refer the reader to Refs.~\cite{A1hydrogen,Silke}.

In this experiment, the argon reached a final temperature of 95~K exiting the nozzle
with a flux of 300~$l_n/h$ ($l_n$ are {\em norm-liters}, the amount of gas equivalent to 1~l at standard conditions).
These thermodynamic parameters were selected to achieve the lowest possible temperature without freezing the argon,
while approaching the liquid phase to increase the density. The flux was set to the maximum value allowed by the vacuum system.
To ensure compatibility with the accelerator vacuum and minimize backgrounds, it was essential to reduce 
the residual argon gas in the scattering chamber. 
Achieving this goal involved testing catchers of various diameters and fine-tuning the distance between the nozzle and the catcher using a remotely controlled step motor.

\begin{figure}
  %\centering
   \resizebox{0.5\textwidth}{!}{%
  \includegraphics{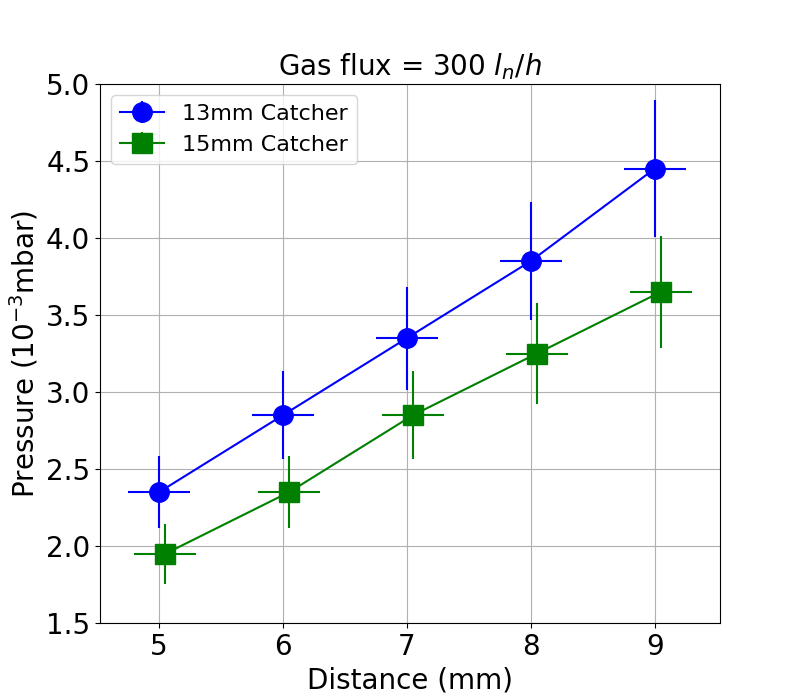}
  }
  \caption{Pressure in the scattering chamber as function of the distance between nozzle and catcher for two
  different catcher diameters and a gas flow rate of 300 $l_n/h$. Error bars reflect the $\pm 10\%$ error
  of the pressure sensor and the $\pm$0.25~mm error in the positioning of the catcher by the step-motor.}
  \label{fig:JTpressure}
\end{figure}

Figure~\ref{fig:JTpressure} shows the pressure in the scattering chamber as a function of the nozzle-to-catcher distance. Tests with two catcher sizes showed a linear relationship between pressure and distance, with the larger catcher achieving a lower pressure in the chamber. 
The distance was not reduced beyond 5~mm to avoid interference with the electron beam and its halo.
The larger catcher, with a 15 mm diameter, combined with the smallest nozzle-to-catcher distance, yielded the best results by achieving the lowest pressure in the scattering chamber.

As shown in Figure~\ref{fig:JTpicture}  (bottom left), the jet is considerably smaller than the catcher's opening. However, a notable difference is observed when using a catcher with a larger diameter. This effect can be attributed to the jet comprising not only a visible central liquid core but also an outer gaseous component, which is optically undetectable and not fully captured by the catcher.

%%%%%%%%%%%%%%%%%%%%%%%%%%%%%%%%%%%%%%%%%%%%%%%%%%%%%%%%%%%%%%%%%%%%%%%%%%%%%%%%%%%%
\section{Estimation of the Target Areal Density}
\label{sec:Operation}

The interaction rate between the electron beam and the jet target is directly dependent on the target's areal density, 
$\rho_{\rm{areal}}$, which must be accurately determined for precise cross section measurements. To achieve this, we first approximate the areal density based on the target's thermodynamic conditions. Analytical equations are utilized for this purpose, providing valuable insights into the influence of various factors contributing to the determination of $\rho_{\rm{areal}}$.

In previous experiments with hydrogen~\cite{A1hydrogen}, the jet shape was measured with a wire scan to be nearly Gaussian, with a width of $\sigma = 1$ mm, measured 5 mm from the nozzle (corresponding to the interaction point with the electron beam). 
In this case, as the jet is clearly visible (see the picture in Figure~\ref{fig:JTpicture}, bottom left), it must be predominantly in the liquid phase and with minimal divergence. 
Since the exact phase composition of the jet is not known, we assume an uniform cylindrical density profile 
and a diameter of $d=2$~mm, corresponding to that of the nozzle.
For this configuration, the areal density can be calculated using the following formula~\cite{A1hydrogen}

\begin{equation}
  %\rho_{areal} = N_{\rm{mol}}\cdot q_V \frac{p_NN_A}{RT_N}\cdot \frac{1}{vA} \cdot 2D \quad,
  \rho_{\rm{areal}} = 4N_{\rm{mol}} \frac{q_V}{\pi dv}\frac{p_N N_A}{T_N R}\,,
  \label{eq:arealdensity1}
\end{equation}
where $v$ is the jet velocity, $q_V$ the flow rate, $N_A$ is the Avogadro number, $N_{\rm{mol}}$ is the number of atoms in the molecule ($N_{\rm{mol}}=1$ for argon), R the universal gas constant, $T_N=273.15$~K the normal 
temperature, and $p_N = 1.01325$~bar the normal pressure.

To estimate the areal density, the velocity of the argon atoms behind the nozzle
must also be known and this depends on the thermodynamic phase of the jet.

For the gaseous state, the velocity can be derived using the ideal gas law~\cite{flux_ideal_gas}
\begin{equation}
  v_{\rm{gas}} = \sqrt{\frac{2\kappa}{\kappa-1} \frac{RT_0}{M}} \,,
  \label{eq:vgas}
\end{equation}
where $\kappa = 5/3$ is the heat capacity ratio for monoatomic gases and $M$ the molecular weight.

In the liquid case, providing an analytical formula for the velocity $v_{\rm{liq}}$
is more difficult and a complex full simulation would be needed. 
%In order to model the gas flow and estimate the gas velocity, 
Therefore, we employ an estimate
assuming  $v_{\rm{gas}} > v_{\rm{liq}}$.
For a liquid with initial pressure $p_0$ flowing through an orifice into the vacuum
we have the lower bound 
\begin{equation}
  v_{\rm{liq}} > \sqrt{\frac{2p_0}{\rho(p_0,T_0)}} \,.
  \label{eq:vliq}
\end{equation}
Using the operational conditions of the jet target with argon ($T_0=95$~K, $p_0=2$~bar with a gas flow $q_N=300$~$l_n/h$)
and $\rho(p_0,T_0)=1438.218$~kg/m$^3$ \cite{NIST} in Eqs.~(\ref{eq:vgas}) and (\ref{eq:vliq}), we obtain $v_{\rm{gas}} =314.4$~m/s and $v_{\rm{liq}} > 16.7$~m/s.

Using Eq.~(\ref{eq:arealdensity1}) with the obtained velocity estimates for the gas and the 
liquid phases we obtain
\begin{eqnarray}
\rho_{\rm{areal}}(\rm{gas})    =  0.46\cdot 10^{18}~\rm{atoms/cm}^2\,, \label{eq:finaldensity1}\\
\rho_{\rm{areal}}(\rm{liquid}) <  8.62\cdot 10^{18}~\rm{atoms/cm}^2 \,.\label{eq:finaldensity2}
\end{eqnarray}
%The target density under the present thermodynamical conditions is expected to be between the latter 
%values.

An additional contribution to the scattering rate is given by the
presence of residual argon
in the scattering chamber that was not fully removed by the catcher. 
We can estimate the areal density of the residual argon $\rho_{\rm{res}}$ with the law of ideal gases corrected by
the spectrometer acceptance $\mathcal{A} = 50$~mm$/\sin\theta_e$ where $\theta_e=20^{\circ}$ is the electron scattering angle
\begin{equation}
  \rho_{\rm{res}} = \frac{p_c}{k_BT_c}\cdot \mathcal{A} = 7.2\cdot 10^{14}~\rm{atoms/cm}^2 \,,
\end{equation}
where $p_c=2\cdot 10^{-3}$~mbar and $T_c=293$~K are the pressure and temperature in the scattering chamber, respectively.
In comparison to the target density, the residual gas contribution can be considered negligible.

%---------------------------
%velocity liquid argon =  16.68 m/s
%velocity gas argon =  314.42 m/s
%Areal density liquid argon =  8.62e+18 /cm2
%Areal density gas argon    =  4.57e+17 /cm2
%Areal density from cross section = 2.9e18 /cm2
%---------------------------
%Acceptance =  14.619022000815438
%Residual volume density (1/cm3) =  4.95E+13 /cm3
%Residual areal density  (1/cm2) =  7.23E+14 /cm2
%---------------------------
%Antoine Eq. Temparature (K) =  24.04 K
%---------------------------

%Stability of the luminosity (lumi monitor)

%%%%%%%%%%%%%%%%%%%%%%%%%%%%%%%%%%%%%%%%%%%%%%%%%%%%%%%%%%%%%%%%%%%%%%%%%%%%%%%%%%%%
\section{Argon Elastic Cross Section Measurement}
\label{sec:CrossSection}
For the first application of the newly developed gas-jet target with argon, we measured the elastic electron scattering cross section, which, in the plane-wave Born approximation (PWBA), is given by the following theoretical expression:
\begin{equation}
    \left( \frac{d\sigma}{d\Omega}\right)_{\!\text{th}} = \left( \frac{d\sigma}{d\Omega} \right)_{\text{\!Mott}} F^2(q^2) \mathcal{R}\quad,
    %\cdot \left( 1 + \frac{2E \sin^2(\theta/2)}{M} \right)^{-1}
    \label{eq:cselastic}
\end{equation}
where
\begin{equation}
        \left( \frac{d\sigma}{d\Omega} \right)_{\text{\!Mott}} = \frac{Z^2 \alpha^2 \cos^2(\theta_e/2)}{4E^2 \sin^4(\theta_e/2)} \quad,
\end{equation}
is the Mott cross section for the scattering of 
a spin-1/2 particle on a spin zero target~\cite{mott}.
$E$ is the incoming electron energy, $Z$ the atomic number,
$\alpha$ the fine structure constant, and $\theta_e$ is the electron scattering angle.
The nuclear form factor $F(q^2)$ depends on the four-momentum transfer squared, $q^2 = -4 E^2 \sin^2(\theta_e/2)$, and
provides information on how protons are distributed within the nucleus.
The factor
\begin{equation}
\mathcal{R} = \left( 1 + \frac{2E \sin^2(\theta_e/2)}{M} \right)^{-1} \quad,
\end{equation}
in Eq.~(\ref{eq:cselastic}) takes into account the recoil of the target nucleus, which
in the present work corrects the cross section only by $\sim 0.1\%$.

To perform the experiment, we use
the 100\% duty cycle electron beam provided by the MAMI accelerator  with an energy $E=705$~MeV and
a nominal current of 3~$\mu$A.
For the measurement itself, two magnetic spectrometers from the A1 experimental setup were used.
Spectrometer B measured the scattered electrons while spectrometer A served as a relative luminosity monitor.
Elastic electron scattering data were recorded at two values of the spectrometer B angle, namely 20$^{\circ}$ 
and 25$^{\circ}$, which correspond to momentum transfers of 1.24~fm$^{-1}$ and 1.55~fm$^{-1}$, respectively.

Due to the large uncertainty in the argon jet density of Eqs.~(\ref{eq:finaldensity1}) and (\ref{eq:finaldensity2}),
the luminosity was determined more precisely by comparison with a numerical calculation that accounts for the interaction of the electron with the protons distributed in the argon nucleus. 

Given that argon has $Z=18$ protons, the PWBA, where the electron is described by a plane wave, does not hold. The distortions of the plane wave due to the nuclear Coulomb field need to be accounted for. This is accomplished using the {\tt ELSEPA} code~\cite{ELSEPA}, which numerically solves the Dirac equation for an electron interacting with a spatially distributed nuclear charge. This approach corresponds to the  so called distorted wave Born approximation (DWBA).

The {\tt ELSEPA} code takes a parameterization of the charge density as input. Here, we use the nuclear charge density 
$\rho_c(r)$ as a function of the distance from the nucleus center $r$, as measured by Ottermann~{\em et al.}~\cite{Ottermann} 
(from 0.59~fm$^{-1}$ to 1.31~fm$^{-1}$ momentum transfers), where it is provided as coefficients $a_i$ from a fit of a 
Fourier-Bessel linear combination to the experimental data
\begin{equation}
  \rho_c(r) = \sum_{i=1}^{N}a_ij_0(q_ir),\quad r<R \,.
  \label{eq:FourierBessel}
\end{equation}
Convergence in the expansion of the above equation is reached with $N=15$ and $\rho_c(r)=0$ if $r\geq R$. The cut-off radius $R=9$~fm was optimized for achieving stability of the fit,
$q_i = \pi i/R$, and $j_0(x)=\sin (x)/x$ is the zeroth-order spherical Bessel function.

The experimental cross section for the $^{40}$Ar(e,e$^{\prime}$) process is defined as
\begin{equation}
  \left(\frac{d\sigma}{d\Omega}\right)_{\rm{exp}} = \frac{N_s-N_b}{\mathcal{L}\cdot\Delta\Omega_{e'}} \frac{1}{\varepsilon} \,,
  \label{eq:cs}
\end{equation}
where $N_s$ and $N_b$ are the number of signal and background events in a scattering angle bin,
respectively, and $\varepsilon$ is the combined efficiency of the detectors. 
$\mathcal{L}$ is the integrated luminosity, and $\Delta\Omega_{e'}$ the angular acceptance,
which was determined with an accurate Monte Carlo simulation of the setup \cite{Miha}.
Background events associated with scattering from the target support structures, nozzle, and catcher 
were estimated by performing measurements with the gas-jet turned off, yielding a result of $N_b = 0$. 

\begin{figure}
    \centering
    \includegraphics[width=1.0\linewidth]{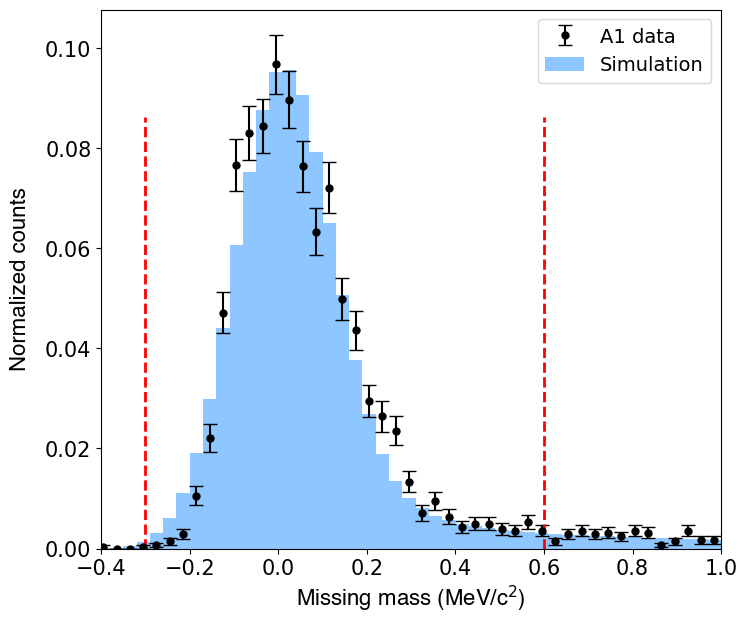}
    \caption{Missing mass peak of the elastic $^{40}$Ar(e,e$^{\prime})$ reaction ($\text{FWHM}=0.24$~MeV). 
    Black points with
    statistical error bars indicates the data, the shaded histogram the simulation.
    The dashed red vertical lines indicate the selection cuts.}
    \label{fig:missingmass}
\end{figure}

Elastic events were selected with cuts on the missing mass $m_{\text{miss}}$ 
which distribution peaks at around $m_{\text{miss}}=0$ for the elastic $^{40}$Ar$(e,e^{\prime})$ reaction (see Fig.~\ref{fig:missingmass}). 
The chosen cuts were -0.3~MeV$< m_{\text{miss}} <$ 0.6~MeV around the missing mass distribution.
The right-hand cut was chosen to include most of the low-energy tail arising from radiative effects, which are well reproduced by the simulation.
The cross-section results remained unchanged when varying the cuts by $\pm 20\%$.

The efficiencies of the scintillation detector, Cherenkov detector,
and vertical drift chambers were determined to be 99\%, 99.85\%, and 99.85\%, respectively, which
are combined multiplicatively into the factor $\varepsilon$.

The integrated luminosity is defined as 
\begin{equation}
  \mathcal{L} = \rho_{\rm{areal}}  \cdot N_e = \rho_{\rm{areal}} \cdot \frac{1}{e}\int_TI(t)dt \,,
  \label{eq:lumi}
\end{equation}
where $N_e$ is the number of incident electrons, $I(t)$ the accelerator current as a function of the time $t$, 
$T$ the total measuring time corrected for the data acquisition dead-time, and $e$ the elementary charge.
The accelerator current was continuously measured non-invasively with a fluxgate magnetometer with $\mathcal{O}$(0.1\%) precision. 

The cross section calculated with {\tt ELSEPA} and the nuclear charge distribution of \cite{Ottermann} was fitted
to the 20$^{\circ}$ dataset using Eq.~(\ref{eq:cs}), leading to the luminosity 
\begin{equation}
\mathcal{L} = 55.1 \pm 2.5 ~ (\mu \rm{b} ~\rm{s})^{-1} \,.
\label{eq:luminosity}
\end{equation}
Using the known beam current, as well as Eq.~(\ref{eq:lumi}) and
Eq.~(\ref{eq:luminosity}), the target
areal density can be inferred, yielding
\begin{equation}
\rho_{\rm{areal}} = (2.9 \pm 0.1)\cdot 10^{18}~\rm{atoms/cm}^2 \,. \nonumber
\end{equation}
The quoted error results from the $\chi^2$ fit of the {\tt ELSEPA}
curve to the data points. 
The obtained areal density shows good consistency with the estimates of Eqs.~(\ref{eq:finaldensity1}) and (\ref{eq:finaldensity2}), which reinforces the reliability of these equations in guiding the target design and optimization.
The same luminosity was applied to the 25$^{\circ}$ dataset, which was acquired under identical experimental conditions.

\begin{figure}
    \centering
    \includegraphics[width=0.965\linewidth]{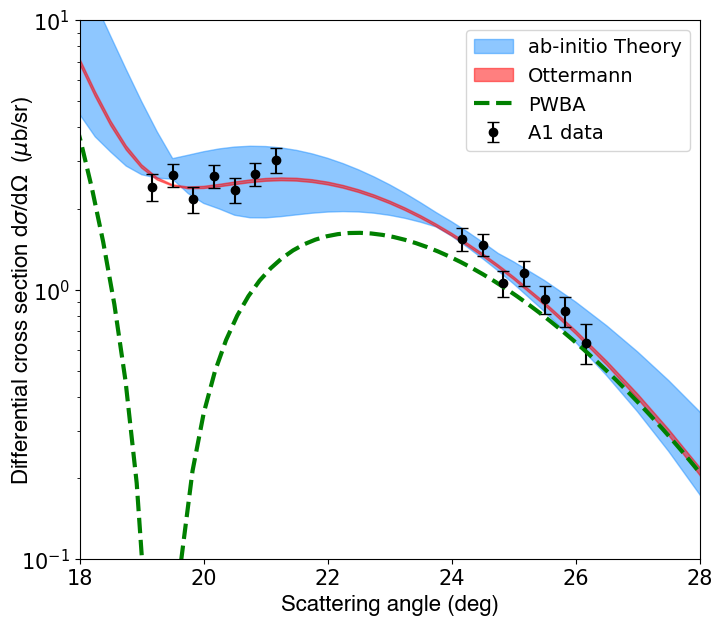}
\caption{Elastic cross section on argon as a function of the electron scattering angle
  at a beam energy of 705~MeV.
    Black points: data from this experiment with statistical error bars. Red band: {\tt ELSEPA} calculation with nuclear charge distribution
    measured in Ref.~\cite{Ottermann}. Blue band: theoretical calculations. Dashed green line: plane wave Born approximation.}
\label{fig:elasticcrosssection}
\end{figure}

Results obtained from this experiment with the gas-jet target are shown in Figure~\ref{fig:elasticcrosssection} 
(black dots). Error bars are purely statistical. 
We compare them to the earlier experimental data by Ottermann {\em et al.}~\cite{Ottermann}, where a high-pressure target (12~bar) was used.
These data are indicated with the red band calculated using {\tt ELSEPA} with the Fourier-Bessel coefficients and the associated statistical errors provided in \cite{Ottermann}. 

%The associated errors are at the order of $\sim 2$\% 
%and are not visible at the scale shown.
Excellent agreement between our new and the old data is observed. This demonstrates the successful operation of the gas-jet target and  validates its capability to measure nuclear cross sections. 

In Figure~\ref{fig:elasticcrosssection}, we also show a comparison to 
{\em ab-initio} calculations,
which are at the forefront of theory in describing properties of light and medium-mass nuclei with quantified uncertainties~\cite{BaccaPastore}.
The presented {\em ab-initio} calculations are based on coupled-cluster theory~\cite{Hagen2014}.
This approach offers a promising framework for modeling particle-nucleus interactions in the context of neutrino searches~\cite{Payne:2019wvy,Sobczyk:2023mey}. 
To explore model uncertainties, we employ Hamiltonians from different  chiral effective field theories at different orders.  
Following~\cite{Payne:2019wvy}, we use two interactions at next-to-next-to-leading order, namely the \NNLOsat~potential~\cite{ekstrom2015} in the $\Delta$-less theory and the \NNLOgod\ potential~\cite{jiang2020} in the $\Delta$-full theory, as well as three different interactions at next-to-next-to-next-to-leading order from the family of Ref.~\cite{Hebeler2011}. %EM1, EM2, EM4 
Different fitting procedures of the low-energy constants are explored: in particular, in the first two Hamiltonians, information on medium-mass nuclei and/or nuclear matter is included, while in the last family of interactions only light nuclei are used, up to $^4$He.
For each Hamiltonian, the charge density of the $^{40}$Ar nucleus is calculated using double-charge exchange coupled-cluster theory~\cite{Payne:2019wvy}, which are then fed to the {\tt ELSEPA} code to obtain a calculation in DWBA. 

The variation of the {\em ab-initio} results due to the different implemented Hamiltonians is shown as an overall (blue) band in Figure~\ref{fig:elasticcrosssection}, which can be seen as an uncertainty to the modeling of the non-perturbative regime of the strong nuclear force.
In the figure, we also show a reference calculation in plane-wave Born approximation (PWBA) as a dashed (green) line, which corresponds to the Fourier transform of the experimental charge density obtained by Ottermann et al.~\cite{Ottermann}. 
Interestingly, the {\em ab-initio} calculations align remarkably well with the shape of the measured cross section at the observed scattering angles. 
The new, relevant information inferred from the results shown in Figure~\ref{fig:elasticcrosssection} is that the distortions accounted for by the {\tt ELSEPA} code reproduce the shape of the experimental data very well, 
filling the dip clearly visible in the PWBA curve, suggesting that the process is modeled very accurately. 

%%%%%%%%%%%%%%%%%%%%%%%%%%%%%%%%%%%%%%%%%%%%%%%%%%%%%%%%%%%%%%%%%%%%%%%%%%%%%%%%%%%%
\section{Summary and Outlook}
\label{sec:Summary}
An argon gas-jet target was used in an electron scattering experiment, enabling the measurement of the nuclear elastic cross section in previously unexplored kinematic regions. 
Operating under thermodynamic conditions close to the liquid state, the target density was enhanced compared to the gas phase.
However, the ill-defined phase of the target made precise calculations of its areal density challenging. Estimates of the areal density, assuming both gaseous and liquid states, were consistent with values obtained by comparing elastic scattering data with a DWBA calculation using a parameterization of the charge density. The result confirms the validity of the
thermodynamical relations and their use in jet target design and optimization.

Finally, we find that our data align closely with modern {\em ab-initio} calculations, highlighting the potential for experiment-theory comparisons to advance our understanding of  the physics of electroweak interactions.

For future operation at MAGIX, the gas-jet target will be complemented with a
dedicated detector for precise online monitoring and absolute determination of the luminosity.
This experiment serves as a proof of concept for utilizing the gas-jet target with elements heavier than hydrogen, 
laying the groundwork for future studies at the MESA facility with the MAGIX experiment. 
By advancing our understanding of electron scattering on argon, this work supports the broader objective 
of enhancing the performance and reliability of argon-based detectors in particle physics.

%%%%%%%%%%%%%%%%%%%%%%%%%%%%%%%%%%%%%%%%%%%%%%%%%%%%%%%%%%%%%%%%%%%%%%%%%%%%%%%%%%%%
\begin{acknowledgement}
The excellent support of the accelerator and technical staff at the Mainz Microtron MAMI is gratefully acknowledged.
L.D. is supported by the Deutsche Forschungsgemeinschaft individual grant No.~521414474 and 
is thankful to X.~Roca-Maza for relevant discussions and explanations about the {\tt ELSEPA} code.
This work is supported by the 
Deutsche Forschungsgemeinschaft project No.~514321794 (CRC1660: Hadrons and Nuclei as discovery tools), 
the US National Science Foundation (NSF) grant PHY-2012114, 
the U.S. Department of Energy, Office of Science, grant number DE-FG02-94ER40818, and
the Research Excellence Initiative of the University of Silesia in Katowice. 
\end{acknowledgement}

%U.S. Department of Energy, Office of Science, grant number DE-FG02-94ER40818

% EPJ FIGURES
%\begin{figure}
%\resizebox{0.75\textwidth}{!}{%
%  \includegraphics{leer.eps}
%\vspace{5cm}       % Give the correct figure height in cm
%\end{figure}
%\begin{figure*}
%\vspace*{5cm}       % Give the correct figure height in cm
%\end{figure*}
%
% For tables use (EPJ)
%\begin{table}
%\begin{tabular}{lll}
%\hline\noalign{\smallskip}
%\end{tabular}
%% Or use
%\vspace*{5cm}  % with the correct table height
%\end{table}
%
% BibTeX users please use
% \bibliographystyle{}
% \bibliography{}
%

\section*{References}
%\begin{thebibliography}{}
\bibliography{bibliography.bib}

\begin{thebibliography}{10}

\bibitem{Rubbia1977}
C.~Rubbia.
\newblock {The Liquid Argon Time Projection Chamber: A New Concept for Neutrino
  Detectors}.
\newblock {\em Report Number: CERN-EP-INT-77-08, CERN-EP-77-08}, 5 1977.

\bibitem{microboone}
R.~Acciarri et~al.
\newblock {Design and construction of the MicroBooNE detector}.
\newblock {\em Journal of Instrumentation}, 12(02):P02017, feb 2017.

\bibitem{ICARUS}
P.~Abratenko et~al.
\newblock Short-baseline neutrino program: initial operation.
\newblock {\em Eur. Phys. J C}, 83:467, 2023.

\bibitem{SBND}
R.~Acciarri et~al.
\newblock {Construction of precision wire readout planes for the Short-Baseline
  Near Detector (SBND)}.
\newblock {\em Journal of Instrumentation}, 15(06):P06033, jun 2020.

\bibitem{DUNE}
B.~Abi et~al.
\newblock {DUNE Far Detector Technical Design Report}.
\newblock {\em JINST}, page T08008, 2020.

\bibitem{DEAP}
Ajaj R. et~al.
\newblock Search for dark matter with a 231-day exposure of liquid argon using
  deap-3600 at snolab.
\newblock {\em Phys. Rev. D}, 100:022004, Jul 2019.

\bibitem{DarkSide50}
Agnes P. et~al.
\newblock {Search for low-mass dark matter WIMPs with 12 ton-day exposure of
  DarkSide-50}.
\newblock {\em Phys. Rev. D}, 107:063001, Mar 2023.

\bibitem{DS20k-0}
C.E. Aalseth et~al.
\newblock Darkside-20k: A 20 tonne two-phase lar tpc for direct dark matter
  detection at lngs.
\newblock {\em Eur. Phys. J. Plus}, 133:131, 2018.

\bibitem{DS20k-1}
P.~Agnes et~al.
\newblock {Sensitivity projections for a dual-phase argon TPC optimized for
  light dark matter searches through the ionization channel}.
\newblock {\em Phys. Rev. D}, 107:112006, Jun 2023.

\bibitem{DS20k-2}
P.~Agnes et~al.
\newblock Sensitivity of future liquid argon dark matter search experiments to
  core-collapse supernova neutrinos.
\newblock {\em JCAP}, 03:043, 2021.

\bibitem{mariani}
H.~Dai et~al.
\newblock {First measurement of the $\mathrm{Ar}(e,e')X$ cross section at
  Jefferson Laboratory}.
\newblock {\em Phys. Rev. C}, 99:054608, May 2019.

\bibitem{e4nu}
A.M. Ankowski et~al.
\newblock Electron scattering and neutrino physics.
\newblock {\em Journal of Physics G: Nuclear and Particle Physics},
  50(12):120501, 2023.

\bibitem{A1hydrogen}
B.S. Schlimme et~al.
\newblock Operation and characterization of a windowless gas jet target in
  high-intensity electron beams.
\newblock {\em Nuclear Instruments and Methods in Physics Research Section A:
  Accelerators, Spectrometers, Detectors and Associated Equipment},
  1013:165668, 2021.

\bibitem{Hofstadter}
R.~Hofstadter.
\newblock Electron scattering and nuclear structure.
\newblock {\em Rev. Mod. Phys.}, 28:214--254, Jul 1956.

\bibitem{MAMI1}
H.~Herminghaus et~al.
\newblock {The design of a cascaded 800 MeV normal conducting C.W. race track
  microtron}.
\newblock {\em Nuclear Instruments and Methods}, 138(1):1--12, 1976.

\bibitem{MAMI2}
K.-H. Kaiser et~al.
\newblock {The 1.5 GeV harmonic double-sided microtron at Mainz University}.
\newblock {\em Nuclear Instruments and Methods in Physics Research Section A:
  Accelerators, Spectrometers, Detectors and Associated Equipment},
  593(3):159--170, 2008.

\bibitem{Magix}
L.~Doria.
\newblock {Electron Scattering for Neutrino Physics at MAMI and MESA}.
\newblock {\em J. Phys.: Conf. Ser.}, 2453:012011, 2023.

\bibitem{MESA1}
Florian Hug, Kurt Aulenbacher, Robert Heine, Ben Ledroit, and Daniel Simon.
\newblock {MESA - an ERL Project for Particle Physics Experiments}.
\newblock 2017.

\bibitem{MESA2}
F.~Hug et~al.
\newblock {Status of the MESA ERL Project}.
\newblock In {\em Proc. ERL'19}, number~63 in ICFA Advanced Beam Dynamics
  Workshop on Energy Recovery Linacs, pages 14--17. JACoW Publishing, Geneva,
  Switzerland, jun 2020.

\bibitem{Epelbaum:2008ga}
Evgeny Epelbaum, Hans-Werner Hammer, and Ulf-G. Meissner.
\newblock {Modern Theory of Nuclear Forces}.
\newblock {\em Rev. Mod. Phys.}, 81:1773--1825, 2009.

\bibitem{machleidt2011}
R.~Machleidt and D.~R. Entem.
\newblock Chiral effective field theory and nuclear forces.
\newblock {\em Phys. Rep.}, 503(1):1 -- 75, 2011.

\bibitem{hammer2020}
H.-W. Hammer, Sebastian K\"onig, and U.~van Kolck.
\newblock Nuclear effective field theory: Status and perspectives.
\newblock {\em Rev. Mod. Phys.}, 92:025004, Jun 2020.

\bibitem{Lovato:2017cux}
A.~Lovato, S.~Gandolfi, J.~Carlson, Ewing Lusk, Steven~C. Pieper, and
  R.~Schiavilla.
\newblock {Quantum Monte Carlo calculation of neutral-current $\nu-^{12}C$
  inclusive quasielastic scattering}.
\newblock {\em Phys. Rev. C}, 97(2):022502, 2018.

\bibitem{Acharya:2019fij}
Bijaya Acharya and Sonia Bacca.
\newblock {Neutrino-deuteron scattering: Uncertainty quantification and new
  $L_{1,A}$ constraints}.
\newblock {\em Phys. Rev. C}, 101(1):015505, 2020.

\bibitem{Payne:2019wvy}
C.~G. Payne, S.~Bacca, G.~Hagen, W.~Jiang, and T.~Papenbrock.
\newblock {Coherent elastic neutrino-nucleus scattering on $^{40}$Ar from first
  principles}.
\newblock {\em Phys. Rev. C}, 100(6):061304, 2019.

\bibitem{Lovato_2020}
A.~Lovato, J.~Carlson, S.~Gandolfi, N.~Rocco, and R.~Schiavilla.
\newblock {Ab Initio Study of
  $({\ensuremath{\nu}}_{\ensuremath{\ell}},{\ensuremath{\ell}}^{\ensuremath{-}})$
  and
  $({\overline{\ensuremath{\nu}}}_{\ensuremath{\ell}},{\ensuremath{\ell}}^{+})$
  Inclusive Scattering in $^{12}\mathrm{C}$: Confronting the MiniBooNE and T2K
  CCQE Data}.
\newblock {\em Phys. Rev. X}, 10:031068, Sep 2020.

\bibitem{Sobczyk:2021dwm}
J.~E. Sobczyk, B.~Acharya, S.~Bacca, and G.~Hagen.
\newblock {Ab initio computation of the longitudinal response function in
  $^{40}$Ca}.
\newblock {\em Phys. Rev. Lett.}, 127(7):072501, 2021.

\bibitem{Sobczyk:2023mey}
Joanna~E. Sobczyk and Sonia Bacca.
\newblock {$^{16}$O spectral function from coupled-cluster theory: Applications
  to lepton-nucleus scattering}.
\newblock {\em Phys. Rev. C}, 109(4):044314, 2024.

\bibitem{HyperK}
K~Abe et~al.
\newblock {Physics potential of a long-baseline neutrino oscillation experiment
  using a J-PARC neutrino beam and Hyper-Kamiokande}.
\newblock {\em Prog. Theor. Exp. Phys.}, page 053C02, 2015.

\bibitem{DEAPeffective}
P.~Adhikari et~al.
\newblock {Constraints on dark matter-nucleon effective couplings in the
  presence of kinematically distinct halo substructures using the DEAP-3600
  detector}.
\newblock {\em Phys. Rev. D}, 102:082001, Oct 2020.

\bibitem{coloma2013}
P.~Coloma and P.~Huber.
\newblock Impact of nuclear effects on the extraction of neutrino oscillation
  parameters.
\newblock {\em Phys. Rev. Lett.}, 111:221802, Nov 2013.

\bibitem{Mihovilovic:2024ymj}
M.~Mihovilovi\v{c} et~al.
\newblock {Measurement of the $\mathrm {{}^{12}C}(e,e')$ Cross Sections at
  $Q^2=0.8\,\textrm{GeV}^2/c^2$}.
\newblock {\em Few Body Syst.}, 65(3):78, 2024.

\bibitem{A1spectrometers}
K.I Blomqvist et~al.
\newblock The three-spectrometer facility at the mainz microtron mami.
\newblock {\em Nuclear Instruments and Methods in Physics Research Section A:
  Accelerators, Spectrometers, Detectors and Associated Equipment},
  403(2):263--301, 1998.

\bibitem{Silke}
S.~Grieser et~al.
\newblock {A cryogenic supersonic jet target for electron scattering
  experiments at MAGIX@MESA and MAMI}.
\newblock {\em Nuclear Instruments and Methods in Physics Research Section A:
  Accelerators, Spectrometers, Detectors and Associated Equipment},
  906:120--126, 2018.

\bibitem{flux_ideal_gas}
A.~T\"aschner, E.~K\"ohler, H.-W. Ortjohann, and A.~Khoukaz.
\newblock {Determination of hydrogen cluster velocities and comparison with
  numerical calculations}.
\newblock {\em The Journal of Chemical Physics}, 139(23):234312, 12 2013.

\bibitem{NIST}
{National Institute of Standards and Technology (NIST)}.
\newblock {NIST Chemistry WebBook, NIST Standard Reference Database Number 69}.
\newblock \url{https://webbook.nist.gov/chemistry}, 2023.

\bibitem{mott}
N.~F. Mott.
\newblock The scattering of fast electrons by atomic nuclei.
\newblock {\em Proceedings of the Royal Society of London. Series A, Containing
  Papers of a Mathematical and Physical Character}, 124(794):425--442, 1929.

\bibitem{ELSEPA}
F.~Salvat, A.~Jablonski, and C.J. Powell.
\newblock {ELSEPA—Dirac partial-wave calculation of elastic scattering of
  electrons and positrons by atoms, positive ions and molecules}.
\newblock {\em Computer Physics Communications}, 165(2):157--190, 2005.

\bibitem{Ottermann}
C.R. Ottermann, CH. Schmitt, G.G. Simon, F.~Borkowski, and V.H. Walther.
\newblock Elastic electron scattering from 40ar.
\newblock {\em Nuclear Physics A}, 379(3):396--406, 1982.

\bibitem{Miha}
M.~Mihovilovič et~al.
\newblock Non-forward radiative corrections to electron-carbon scattering.
\newblock {\em Eur. Phys. J. A}, 59:225, 2023.

\bibitem{BaccaPastore}
Sonia Bacca and Saori Pastore.
\newblock Electromagnetic reactions on light nuclei.
\newblock {\em Journal of Physics G: Nuclear and Particle Physics},
  41(12):123002, 2014.

\bibitem{Hagen2014}
G.~Hagen, T.~Papenbrock, M.~Hjorth-Jensen, and D.~J. Dean.
\newblock Coupled-cluster computations of atomic nuclei.
\newblock {\em Reports on Progress in Physics}, 77(9):096302, 2014.

\bibitem{ekstrom2015}
A.~Ekstr\"om, G.~R. Jansen, K.~A. Wendt, G.~Hagen, T.~Papenbrock, B.~D.
  Carlsson, C.~Forss\'en, M.~Hjorth-Jensen, P.~Navr\'atil, and W.~Nazarewicz.
\newblock Accurate nuclear radii and binding energies from a chiral
  interaction.
\newblock {\em Phys. Rev. C}, 91:051301, May 2015.

\bibitem{jiang2020}
W.~G. Jiang, A.~Ekstr\"om, C.~Forss\'en, G.~Hagen, G.~R. Jansen, and
  T.~Papenbrock.
\newblock Accurate bulk properties of nuclei from $a=2$ to
  $\ensuremath{\infty}$ from potentials with $\mathrm{\ensuremath{\Delta}}$
  isobars.
\newblock {\em Phys. Rev. C}, 102:054301, Nov 2020.

\bibitem{Hebeler2011}
K.~Hebeler, S.~K. Bogner, R.~J. Furnstahl, A.~Nogga, and A.~Schwenk.
\newblock Improved nuclear matter calculations from chiral low-momentum
  interactions.
\newblock {\em Phys. Rev. C}, 83:031301, Mar 2011.

\end{thebibliography}
\bibliographystyle{unsrt}
%\end{thebibliography}

%\begin{thebibliography}{}

%Author, \textit{Book title} (Publisher, place year) page numbers
% etc

%\end{thebibliography}

\end{document}